\documentclass[11pt]{article}

\usepackage{amsmath,amsfonts,amssymb,graphicx,geometry,authblk,xcolor,subfigure} 
\usepackage{gensymb}
\usepackage[title]{appendix}
\newcommand{\br}{x} 
\newcommand{\bk}{k} 

\title{Machine-learned prediction of the electronic fields in a crystal}
\author[1]{Ying Shi Teh}
\author[2]{Swarnava Ghosh}
\author[1]{Kaushik Bhattacharya}

\affil[1]{Division of Engineering and Applied Science, California Institute of Technology, Pasadena CA 91125}
\affil[2]{National Center for Computational Sciences, Oak Ridge National Laboratory, Oak Ridge, TN 37830 } 

\date{}

\begin{document}

	\maketitle

\begin{abstract}
We propose an approach for exploiting machine learning to approximate electronic fields in crystalline solids subjected to deformation.  Strain engineering is emerging as a widely used method for tuning the properties of materials, and this requires repeated density functional theory calculations of the unit cell subjected to strain.  Repeated unit cell calculations are also required for multi-resolution studies of defects in crystalline solids.  We propose an approach that uses data from such calculations to train a carefully architected machine learning approximation.  We demonstrate the approach on magnesium, a promising light-weight structural material: we show that we can predict the energy and electronic fields to the level of chemical accuracy, and even capture lattice instabilities.

\end{abstract}

\section{Introduction}

A number of studies over the recent years have shown that the electronic structure of crystalline solids depends sensitively on the deformation, and therefore straining a lattice from its equilibrium structure can lead to new properties.  For example, the perovskite SrTiO$_3$ which is usually paraelectric becomes ferroelectric when subjected to a lattice strain \cite{setal_armr_03}.   Silicon becomes electrically polarized under strain, and the role of strain on various functional materials has been extensively studied \cite{tetal_aip_19}.  In metals, strain can lead to deformation twinning \cite{cm_progmatsci_95}.  Strain-induced martensitic phase transitions are widely observed and exploited in shape-memory alloys and steels \cite{Lu1997,Iwamoto1998,Qin2020}.  Finally, strain engineering is emerging as an important tool in 2D materials \cite{Dai2019}.   

Density functional theory (DFT) \cite{Kohn1965} is a powerful tool to understand the electronic structure of matter, and thus has been instrumental to the understanding, design, and optimization of materials.  Examples include the predictions of energy materials \cite{Jain2016}, the geometric design of polar metals \cite{Kim2016}, and the screening for high-performance piezoelectrics \cite{Armiento2011}.  
Strain-induced phenomena can also be studied using DFT but it requires the repeated electronic structure calculation of a crystalline lattice unit cell subject to various strains.  Consequently, a systematic exploration of the strain space  can be computationally expensive.  In this work, we study if a neural network approximation can assist in this exploration.  
We are motivated by the success of machine learning, particularly deep neural networks in image recognition \cite{lecun1995convolutional, he2016deep} and natural language processing tasks \cite{goldberg2017neural, collobert2008unified}.   There is also a growing literature on the use of these methods in materials science \cite{kd_armr_15}.   

Another motivation for our work comes from the study of defects in crystalline solids that play a critical role in determining mechanical and other properties of various solids: for example, vacancies are critical for creep and dislocations for plasticity.  The perturbations caused by these defects decay very slowly, and therefore their direct study requires very large computational domains.   Solving the DFT equations is prohibitive on such large computational domains, and a variety of approaches have been proposed (e.g., quantum mechanics/molecular mechanics \cite{Ogata2001,Csanyi2004,Khare2007,Kermode2008,Bernstein2009} and embedded DFT \cite{wetal_prl_08}).
Fago \textit{et al.} \cite{Fago2004} introduced a ``local'' DFT-based quasicontinuum method where the deformation of the atoms is assumed to follow a piece-wise affine deformation, and the energy density of each region is computed using a unit cell DFT calculation.
A new approach was introduced by Suryanarayana \textit{et al.} \cite{Suryanarayana2012} that solves the DFT equations by introducing a numerical basis that exploits the decay.   Specifically, the electronic fields are taken to be a sum of a piece-wise periodic `predictor' and a slowly decaying `corrector.'  The approach leads to accurate solutions over millions of atoms. It can resolve  the core, the far field, as well as the interactions between far-field stress and core of defects including dislocations \cite{Ponga2016,pbo_jcp_20}.  The implementation of these approaches also requires the repeated solution of the unit cell subject to distortion.

In this paper, we study a deep neural network approximation for the  energy and the underlying electronic fields in a unit cell of magnesium subjected to strain.  We generate data by repeatedly solving the unit cell problem and use it to train a deep neural network.    An important challenge is the representation of the electronic fields: these are elements of infinite dimensional function spaces whereas neural networks typically approximate maps between finite dimensional spaces.  Therefore, we use the approach that combines model reduction and neural networks for high-fidelity approximations of maps between function spaces \cite{Bhattacharya2020}.   We show excellent, specifically chemical, accuracy of the trained neural network approximation over a range of strain.  Further, the approximation is able to learn the onset of an instability.

Some recent works have focused on approximating electronic structure quantities, mainly electronic density, using machine learning as a means to bypassing DFT calculations. Chandrasekaran \textit{et al.} \cite{Chandrasekaran2019} used a representation that encodes the atomic arrangement around any grid point and mapped it to electron charge density and local density of states (LDOS) spectrum at the corresponding grid point. This grid point method allows the quantities at each grid point to be evaluated independently and hence in parallelization. It can also be highly dependent on the discretization used and quickly become intractable as the system size grows. Grisafi and co-workers \cite{Grisafi2019} expanded the electronic charge densities of different hydrocarbons as sums of atom-centered basis functions and machine-learned them using symmetry-adapted Gaussian process regression. The use of such localized basis set allows for transferability across different molecular systems, but is not applicable to metals. A work by Brockherde \textit{et al.} \cite{Brockherde2017} explores the mappings from potential to ground-state density represented in Fourier basis. One interesting finding is that learning energy indirectly -- from potential to electron density followed by electron density to energy -- yields better predictions than the direct map from potential to energy. It is noteworthy that all these methods use the atomic environment typically within some cutoff radius as the input of the learning maps, and this requires careful selection of descriptors. Our work focuses on crystals or solids and uses strain as its input. Its simplicity allows for highly accurate predictions of electronic fields along with the combination of data-based model reduction and neural networks. It also offers a convenient way to relate electronic structure calculations to  larger continuum level constitutive response of materials.

\section{Background}\label{sec:background}

\subsection{Density functional theory}

Given atomic positions $\{R_I\}$, density functional theory seeks to find the total electronic free energy  system $\mathcal{F}(\{R_I\})$, the electronic charge density $\rho(x)$, and other electronic functions of interest.  To do so, we solve the Kohn-Sham equation for energy states $E_i$ and orbitals $\psi_i(x)$ (ignoring spins for simplicity of presentation and assuming a non-local pseudo-potential in the Kleinman-Bylander form),
\begin{equation} \label{eq:ks}
\mathcal{H} \psi_i = E_i \psi_i, \quad 
\mathcal{H} = - \frac{1}{2}\nabla^2 + V^\text{nl}_\text{ps} + V_\text{xc} + V_\text{H} + V_\text{ext} 
\end{equation}
where $V^\text{nl}_\text{ps}(x,x'; \{R_I\}) $ is  the non-local portion of the psuedopotential and depends on the atomic positions $\{R_I\}$, and $V_\text{xc}, V_\text{H}, V_\text{ext}$ are the exchange-correlation, Hartree, and external potential due to ions given by 
\begin{equation}
V_\text{xc} = {\partial e_\text{xc} \over \partial \rho}, \quad - \frac{1}{4\pi} \nabla^2 V_\text{H} = \rho, \quad - \frac{1}{4\pi} \nabla^2 V_\text{ext} = b,
\end{equation}
with valence electron density $\rho$, charge density $b = b(x; \{R_I\})$ describing the local part of the pseudopotential, and exchange-correlation density $e_{xc} (\rho) $.   It is convenient to write the Hamiltonian $\mathcal{H}$ in operator form, and introduce the corresponding (one-point) density operator whose diagonal component gives the electron density \begin{equation} 
\mathcal{H} = \sum_i E_i \psi_i(x) \psi_i(x'), \quad 
\gamma(x,x') =  \sum_i f(E_i) \psi_i(x) \psi_i(x'), \quad
\rho(x) = \gamma(x,x),
\end{equation}
where $f$ describes the occupancy and satisfies $\sum_i f(E_i) = n$ with $n$ being the total number of electrons in the system.  The total electronic free energy of the system may be written as
\begin{equation} \label{eq:free}
\mathcal{F}(\{R_I\}) = \text{Tr }(\mathcal{H} \gamma) + \int \left( e_\text{xc} - V_\text{xc} \rho - {1\over 2}  V_\text{H} \rho + {1\over 2}  V_\text{ext} b \right) dx - \text{Tr} (S(\gamma))/\beta,
\end{equation}
where $\beta=1/(k_B T)$, $T$ is the fictitious electronic temperature, and $S$ is the generalized entropy that determines the occupancy $f$.  
We label the first term of (\ref{eq:free})  the band structure energy  ( $U = \text{Tr}(\mathcal{H} \gamma)$) and note that its density is
\begin{equation}
u(x) =  \sum_i E_i f(E_i) |\psi_i(x)|^2.
\end{equation}

If we view $\mathcal F$ as a functional on $\gamma$, then the Kohn-Sham equation (\ref{eq:ks}) and an equation for $f$ are the Euler-Lagrange equation associated with the variational problem.  Note that the Kohn-Sham equation is non-linear because the exchange-correlation and the Hartree potential depend on the electron density, so it is usually solved by fixed point iteration (also known as the self-consistent field approach).  

We specifically consider the entropy associated with the Marzari cold smearing with broadening \cite{Marzari1996} 
\begin{equation}
{ \frac{1}{\beta} f'(E)}  = \frac{2}{\sqrt{\pi}} \left( \kappa t^3 - t^2 - \frac{3}{2} \kappa t + \frac{3}{2} \right) e^{-t^2}, \quad t= \beta(E - E_f) .
\end{equation}
$\kappa = -0.5634$ and $E_f$ is the Fermi energy or the Lagrange multiplier that enforces the constraint $\sum_i f(E_i) = n$\footnote{We have also repeated our work with the entropy of mixing $S(f) = - f \log {f\over2} + (2-{f\over2}) \log (1-{f\over2})$ that gives $f\over2$ to be the Fermi-Dirac function.}. The volumetric entropy (i.e. entropy per unit volume) is
\begin{equation}
s(x) = -\frac{1}{\sqrt{\pi}} \sum_i |\psi_i(x)|^2 e^{-t^2} (\kappa t^3 + t^2 - \frac{1}{2}).
\end{equation}
\vspace{0.1in}

Putting all these together, we view density functional theory as a map
\begin{equation} 
\Phi_\text{DFT}: \{R_I\} \to \{ \rho(x), \phi(x), u(x), s(x), \mathcal{F}\},
\end{equation}
where $\phi = V_\text{H}+V_\text{ext}$ is the Coulomb potential.

\subsection{Crystals}

A crystal is a periodic arrangement of $N$ atoms described by a unit cell $\mathcal{U}$ bounded by three lattice vectors $\{ a, b, c \}$ and $N$ atomic positions or basis vectors $R_I, I = 1, \dots, N$.  It is customary to introduce fractional coordinates $\bar{R}_{I}$ with respect to the lattice vectors.

Using the Bloch theorem, the electronic orbitals may be written as $\psi_{i,k} = \exp(i k \cdot x) \Psi_{i,k} (x)$ where $\Psi_{i,k}$ is periodic and $k$ is a vector in the Brillouin zone associated with $\mathcal{U}$.  The formulas above can be naturally extended (see Appendix \ref{appendix:derivation}) and we obtain $\rho(x), \phi(x), u(x), s(x),$ to be periodic functions while the free energy $\mathcal{F}$ is now interpreted as energy per unit cell.

We are interested in the deformations of the crystal,  so we choose a reference crystal structure with lattice vectors  $\{ a^0, b^0, c^0 \}$ and atomic coordinates $\{\bar{R}_I^0\}$.  We can then describe the deformation (up to rotations) in terms of  
\begin{equation}
D = \{\lambda_a, \lambda_b, \lambda_c, \theta_a, \theta_b, \theta_c\} \quad \text{where} \quad
\lambda_a  = {|a| \over |a^0|},  \ \theta_a = \arcsin\left( {b \times c \over |b| |c|} \right), \text{ etc.}
\end{equation}

Now, given any deformed crystal and any set of atomic coordinates, we can find the electronic states by solving the electronic states as described above.  Further, we can find the equilibrium states of the atoms $\{\bar{R}_I^e\}$ by solving ${\partial \mathcal{F} \over \partial \bar{R}_I} = 0, I = 1, \dots N-1$.  

Finally, the electronic quantities are functions defined on the deformed unit cell or a domain that depends on the strain.  It is convenient to define them on a fixed domain,  so we map them back to the reference lattice with a change of variables $\bar{\rho}(F^{-1}x) = \rho(x)$, $\bar{\phi}(F^{-1}x) = \phi(x)$, etc., where $F$ is a tensor that maps the reference unit cell to the deformed unit cell $a = Fa^0, b= Fb^0, c= Fc^0$.  

In summary, the deformation behavior of a crystal is described by the map
\begin{equation} 
\Phi: D \to \{ \{\bar{R}_I^e\}, \bar \rho(x), \bar \phi(x), \bar u(x), \bar s(x), \mathcal{F}\},
\end{equation}
where the electronic states are computed for the deformed crystal with the atoms in their equilibrium positions.

\subsection{Implementation}

The density functional theory calculations to evaluate the map $\Phi$ are conducted using the software ABINIT \cite{Gonze2020}.  We use a plane-wave basis set with a kinetic energy cut-off of 24 Ha (Hartree), a Troullier-Martins norm-conserving pseudopotential with local channel $l=1$, and local density approximation (LDA) in the Perdew-Wang 92 functional form as the exchange-correlation energy.   Cold smearing of magnitude 0.01 Ha is used \cite{Marzari1996} and the Brillouin zone integration is performed using a $12\!\times\!12\!\times\!12$ k-point sampling. Furthermore, the atomic positions are relaxed using the Broyden-Fletcher-Goldfarb-Shanno minimization.

\section{Approach}\label{sec:approach}

We seek to learn an approximation for the map $\Phi$ for a given material.  We first generate data by evaluating the map using DFT and seek to use this data to learn an approximation.  However, note that the quantities $\rho, \phi, u, s$ are functions and thus elements of infinite-dimensional linear spaces.  In practice, these are evaluated on a finite-dimensional discretization, but still we want our approximation to be independent of the particular discretization.  Therefore we use an approach by Bhattacharya {\it et al.} \cite{Bhattacharya2020} that combines model reduction with a deep neural net to learn the map $\Phi$.  The idea is to use model reduction to find a finite dimensional representation for each function and then use a deep neural net to learn the map.  Specifically, we find maps $p_\rho: \bar \rho \to \{\rho_\alpha\}_{\alpha=1}^{d_\rho}$, $p_\phi: \bar \phi \to \{\phi_\alpha\}_{\alpha=1}^{d_\phi}$, etc.\ that reduce (project) the infinite-dimensional spaces to a $d_\rho$-dimensional space, and maps $\ell_\rho:  \{\rho_\alpha\}_{\alpha=1}^{d_\rho} \to \bar \rho$, $\ell_\phi:  \{\phi_\alpha\}_{\alpha=1}^{d_\phi} \to \bar \phi$, etc.\ that lift (reconstruct) the 
$d_\rho$-dimensional space to the infinite dimensional space.  We then find an approximate map  
\begin{equation}
\Phi_\text{ml}: D \to \{ \{\bar{R}_I^e\}, \{ \rho_\alpha \}, \{\phi_\alpha\}, \{u_\alpha\}, \{s_\alpha\}, \mathcal{F}\}
\end{equation}
such that $\Phi \approx \ell \circ \Phi_\text{ml} $.
In this work, we use principal component approximation (PCA) for model reduction $p, \ell$ and a deep neural net for $\Phi_\text{ml}$.

\section{Demonstration on Magnesium}\label{sec:mg}

\subsection{Magnesium}

Magnesium is a hexagonal close-packed (HCP) material. It is the lightest of all structural materials, and of significant interest as a light-weight structural material for bio-medical, automotive, and protective applications \cite{Joost2017,Kulekci2008,Chen2016}.  
\begin{figure}
	\begin{center}
		\includegraphics[width=0.5\textwidth]{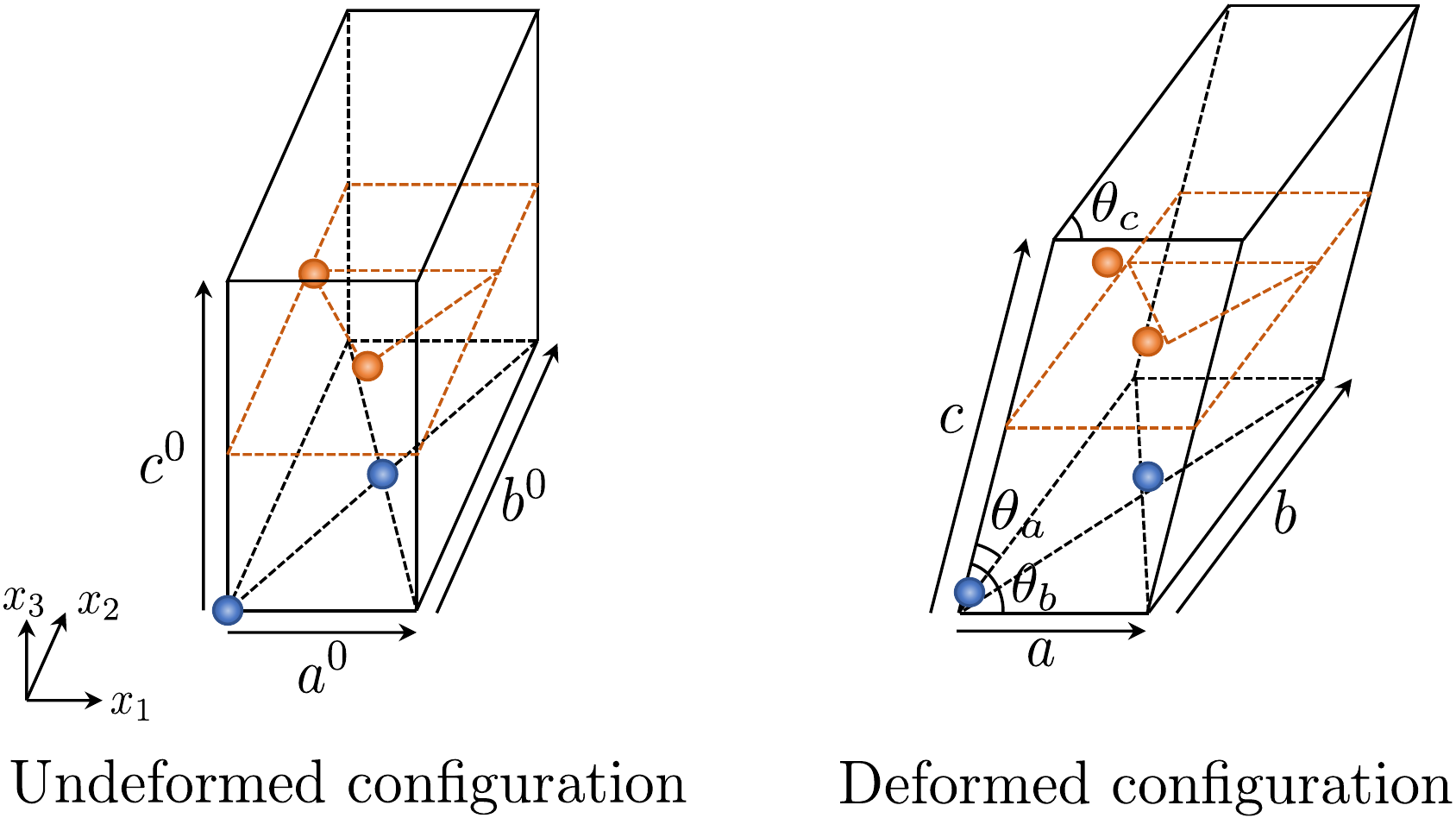}
	\end{center}
	\caption{A 4-atom magnesium unit cell in both undeformed and deformed configurations.}
	\label{fig:Mg unit cell}
\end{figure}
We consider the four-atom unit cell shown in Figure \ref{fig:Mg unit cell} where $a^0$, $b^0$, and $c^0$ directions are the $[10\bar{1}0]$, $[0,\bar{1},1,0]$, and $[0001]$ directions, respectively, in the HCP crystallographic notation.   The atoms in the reference unit cell are located at fractional coordinates $\{0,0,0\}$, $\{1/2,1/2,0\}$, $\{0,2/3,1/2\}$, and $\{1/2,1/6,1/2\}$ in the reference configuration.  The $a^0$-$b^0$ plane perpendicular to the $c^0$-axis or $(0001)$ plane is the basal plane.  We have observed in our calculations that the basal planes deform uniformly, but slide relative to each other. After eliminating free translation of the unit cell, the fractional coordinates in the deformed configuration can be taken as $\{\bar{R}_1^e,\bar{R}_2^e,\bar{R}_3^e\}, \{1/2+\bar{R}_1^e,1/2+\bar{R}_2^e,\bar{R}_3^e\}, \{-\bar{R}_1^e,2/3-\bar{R}_2^e,1/2-\bar{R}_3^e\}, \{1/2-\bar{R}_1^e,1/6-\bar{R}_2^e,1/2-\bar{R}_3^e\}$.

\subsection{Architecture and training}

We use PCA dimensions of $d_\rho \!=\! d_\phi \!=\! d_u \!=\! d_s \!=\! 50$ and the following neural network architecture: (1) a two-layer dense network with hidden layer widths of 500 and the hyperbolic tangent activation function for each of $\Phi_\text{ml}^\rho$,$\Phi_\text{ml}^\phi$,$\Phi_\text{ml}^u$ and $\Phi_\text{ml}^s$; (2) a three-layer dense network with hidden layer widths of 50, 100, 50, respectively, and the same type of activation function for $\Phi_\text{ml}^R$ and $\Phi_\text{ml}^\mathcal{F}$. These hyperparameters are selected based on four-fold cross-validation results. 

We generate a total of 3000 data, with each input sample $D$ drawn independently from a normal distribution truncated to two standard deviations satisfying $\lambda_a,\lambda_b,\lambda_c \in [0.9,1.1]$ and $\theta_a,\theta_b,\theta_c \in [84\degree,96\degree]$. Such a distribution reflects the fact that smaller deformations are more likely to be encountered in real materials. Out of all the data generated, 2000 of them are reserved for training and validation, while the rest are used for testing.

Using the training data, we first identify the map $p$ via PCA. This is followed by standardizing both the input and output of $\Phi_{\text{ml}}$ to zero mean and unit variance, before we train the neural network parameters using the Adam optimization algorithm at a training rate of 0.001, a small l2 regularization of 0.0001 on the weights, and a batch size of 128 for a total of 4000 epochs. Subsequently, given any deformation in the testing data, we generate predictions by applying the map $\ell \circ \Phi_\text{ml} $ and compare them with the true values.

\subsection{Computational costs}

There are two elements to the computational cost.  The first is the online cost of evaluation.  This takes fractions of a second (0.002 second) on an Intel Skylake (2.1GHz) core compared to 30 minutes on 14 cores  for a full DFT evaluation.  Thus, learned approximations provide significant savings.  The second is the one-time offline cost of generating the data and training.   As noted, each data set takes 30 minutes on 14 cores Intel Skylake (2.1GHz) and we generate 2000 data sets for training.  This is comparable to a single evaluation in a MacroDFT calculation.  However, since each data set is independent, it is is trivially parallelizable.  The cost of training is about 5 minutes on a single core.
%

\subsection{Results: Electronic fields and energy}

\begin{figure}[t]
	\begin{center}
		\includegraphics[width=1.0\textwidth]{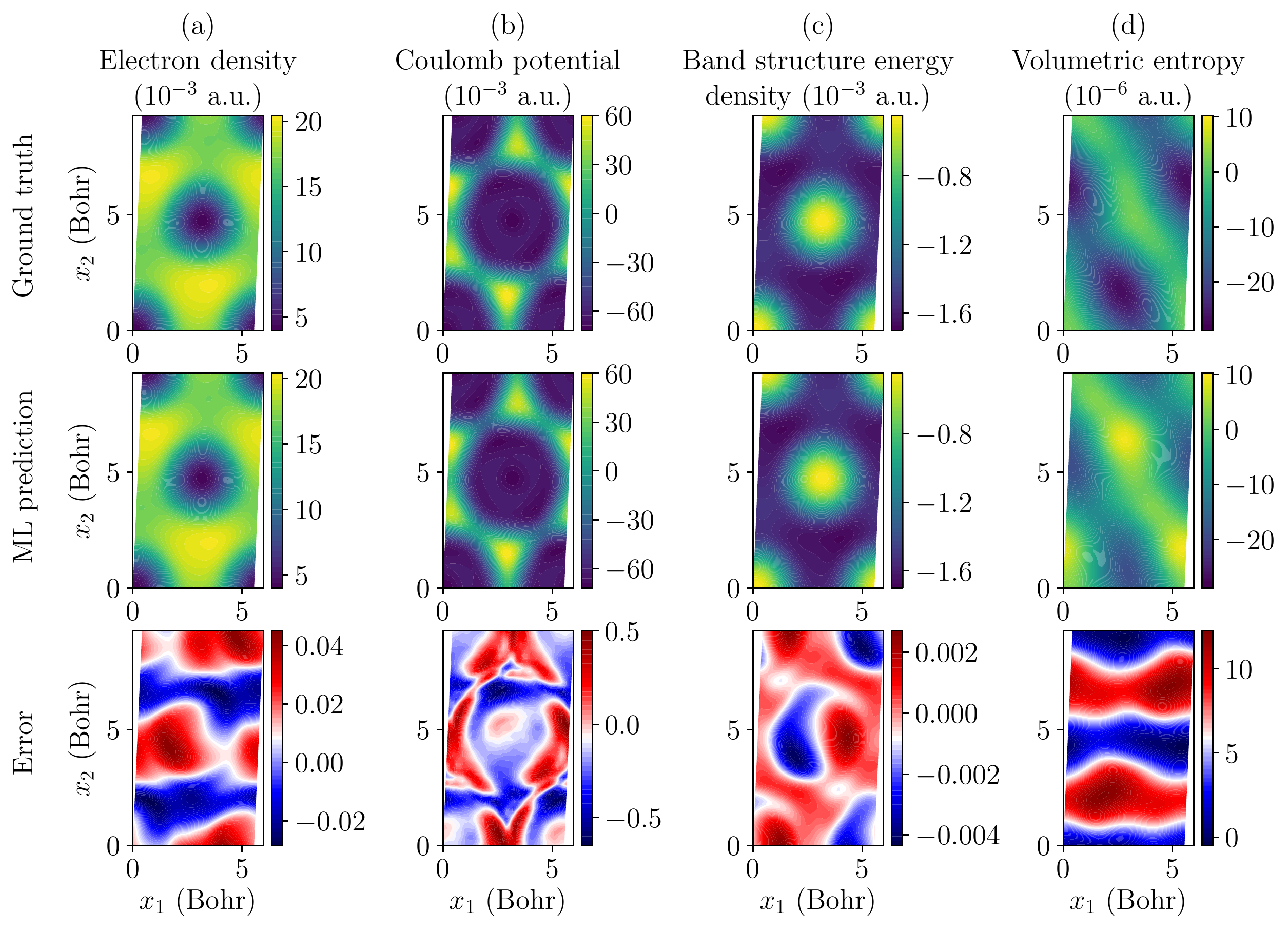}
	\end{center}
	\caption{Typical results.  (a) Electron density, (b) Coulomb potential, (c) band structure energy density, and (d) volumetric entropy in atomic unit (a.u.) along the $x_3=0$ plane for  $\lambda_a = 0.9696, \lambda_b = 0.9237, \lambda_c = 0.9906, \theta_a = 92.1598\degree, \theta_b = 85.6196\degree, \theta_c = 87.3824\degree$.  Notice that the scale used to display the error is significantly smaller than the scale used to display the quantities except for entropy, which is small.}
	\label{fig:prediction sample}
\end{figure}

\begin{figure}[t]
	\begin{center}
		\includegraphics[width=0.9\textwidth]{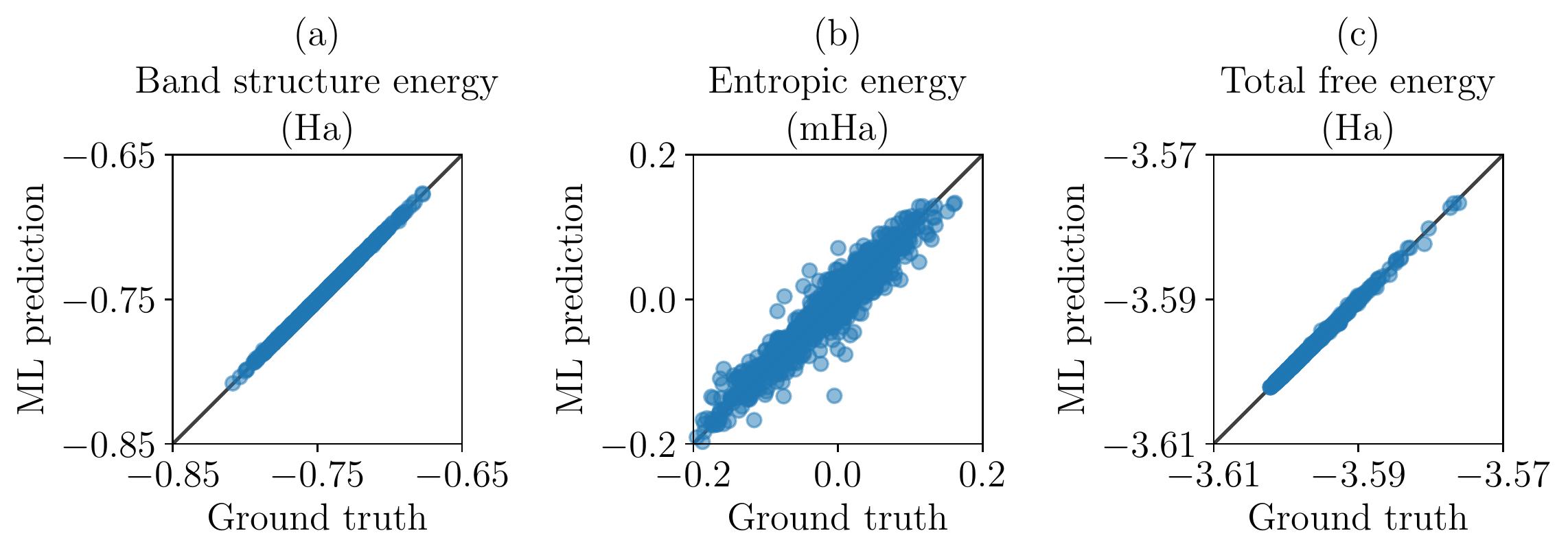} 
	\end{center}
	\caption{Comparisons between predicted and true values for (a) band structure energy ($U=\int u(x) dx$), (b) entropic energy ($-S/\beta = - \int s(x) dx / \beta$), and (c) total free energy ($\mathcal{F}$). Predictions are perfectly accurate if all the data points lie on the black solid line of $y=x$. The mean errors are 0.15 mHa, 0.014 mHa, and 0.10 mHa while the maximum errors are 2.4 mHa, 0.13 mHa, and 1.5 mHa for (a), (b) and (c), respectively.}
	\label{fig:prediction energy}
\end{figure}

A typical result is shown in Figure \ref{fig:prediction sample}.  We observe that our approach is able to capture the main features of the electronic fields, with very small errors.  Figure \ref{fig:prediction energy} compares the predicted and actual energies (band structure energy, entropic energy, and total free energy).  The mean errors are 0.15 mHa, 0.014 mHa, and 0.10 mHa while the maximum errors are 2.4 mHa, 0.13 mHa, and 1.5 mHa for Figure \ref{fig:prediction energy} (a), (b), and (c), respectively.  
Importantly, since there are four atoms in this unit cell, all the errors are significantly smaller than 1.6 mHa/atom (or 1 kcal/mol) which is widely accepted as the accuracy required for chemical accuracy \cite{Pople1999}.

\begin{figure}[t]
	\begin{center}
		\includegraphics[width=1.0\textwidth]{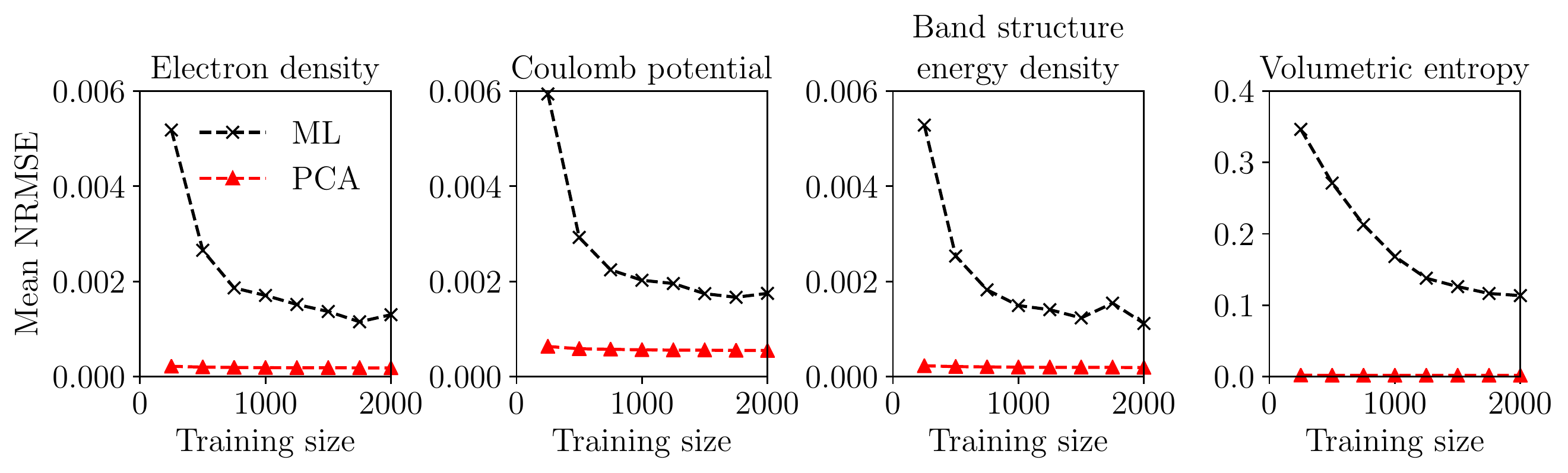}
	\end{center}
	\caption{Training error.  Variation of test error (NRMSE as in equation (\ref{eqn:NRMSE})) marked as ML with training size for the four scalar field quantities. The NRMSE shown in the plot is averaged over all 1000 test samples.  The figure also shows the PCA error.}
	\label{fig:trainingerror}
\end{figure}

\begin{figure}[t]
	\begin{center}
 		\includegraphics[width=1.0\textwidth]{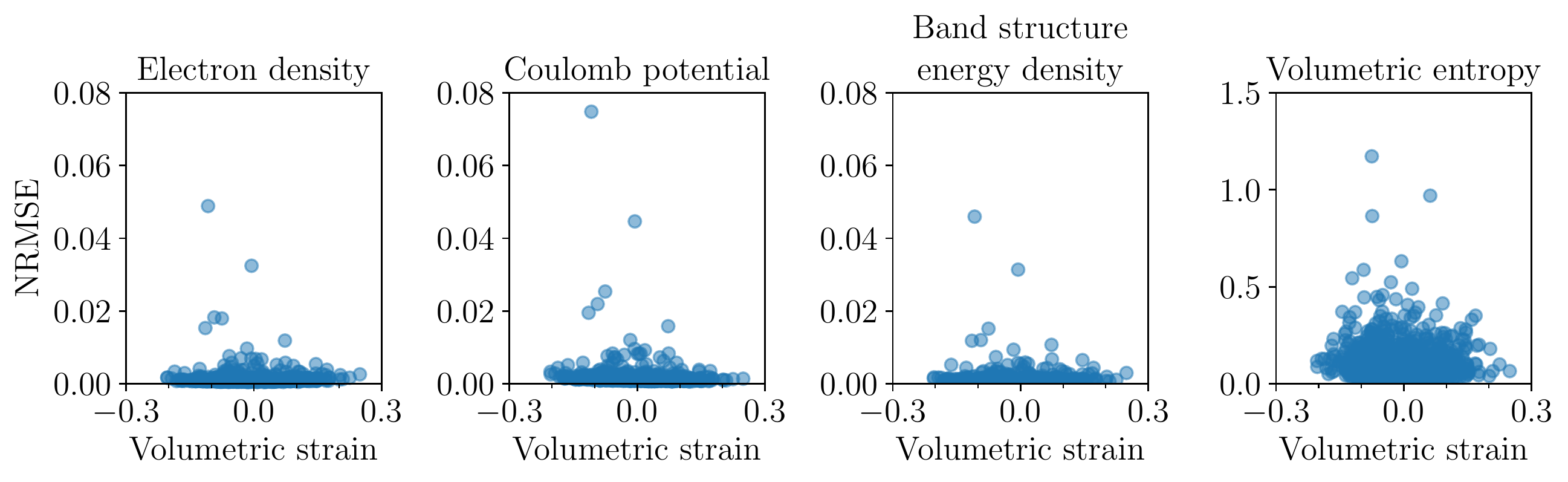}
	\end{center}
	\caption{Distributions of normalized root-mean-square errors (NRMSE) between predicted and true values evaluated on a $36\times64\times60$ grid across 1000 test samples.}
	\label{fig:error vs volumetric strain}
\end{figure}

We now turn to understanding the training and the actual distribution of errors.   We introduce a normalized root-mean-square error (NRMSE), defined as root-mean-square error divided by the range of the data (evaluated on a discretized grid of size $N_d$ for field quantities)
\begin{equation}\label{eqn:NRMSE}
\text{NRMSE} = \frac{\sqrt{ \frac{1}{N_d}\sum_{j=1}^{N_d}(y_j^{pred}-y_j^{true})^2 }}{y_{max}^{true} - y_{min}^{true}}.
\end{equation}
Figure \ref{fig:trainingerror} shows how the normalized root-mean-square error (evaluated on a $N_d=36\times64\times60$ grid for field quantities) averaged over the 1000 test samples decreases with an increasing number of training samples.   We see that the error has stabilized at 2000 training samples for this set of test samples.   The figure also shows the error due to PCA, i.e., the error associated with the model reduction from an infinite dimensional function space to a finite dimensional representation of the electronic fields.  We see that the PCA error is about ten times smaller than the overall error.     Figure \ref{fig:error vs volumetric strain} shows how the NRMSE changes with volumetric strain for these 1000 test samples.  We notice that the error in electron density, Coulomb potential, and band structure energy density is extremely small ($\sim$0.1\%) in almost all cases with a maximum error of $\sim$8\% in isolated cases.   Indeed, the five points with the largest error in all these plots all correspond to the same five test data points and  are  associated with  very large distortions in the unit cell that were poorly represented in the normal distribution used to sample the training data. 
The error is larger in volumetric entropy averaging $\sim$11\% mainly because this is a very small number.
Finally, the error displacement of the basis atoms is 0.0035 Bohr which is much smaller than the maximum and average atomic displacements of 0.66 Bohr and 0.097 Bohr, respectively. Again, the five points with the largest error here correspond to the same data points with large error in Figure \ref{fig:error vs volumetric strain}.

\begin{figure}[t]
	\begin{center}
		\includegraphics[width=0.7\textwidth]{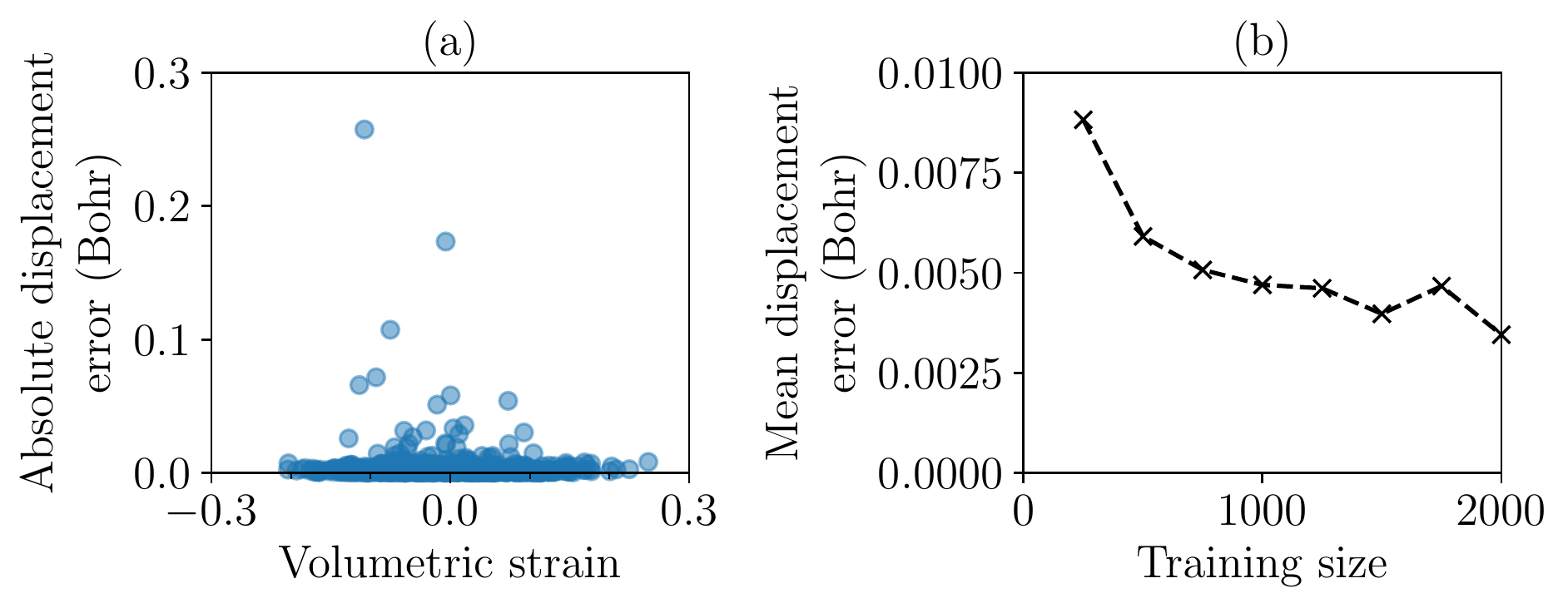}
	\end{center}
	\caption{Test errors for reduced coordinates of magnesium atoms in terms of absolute displacement errors. (a) shows the distribution of errors for 1000 test samples. (b) shows how the mean error varies with training size.}
	\label{fig:XRED displacement error vs volumetric strain}
\end{figure}

\begin{figure}[t]
	\begin{center}
		\includegraphics[width=1.0\textwidth]{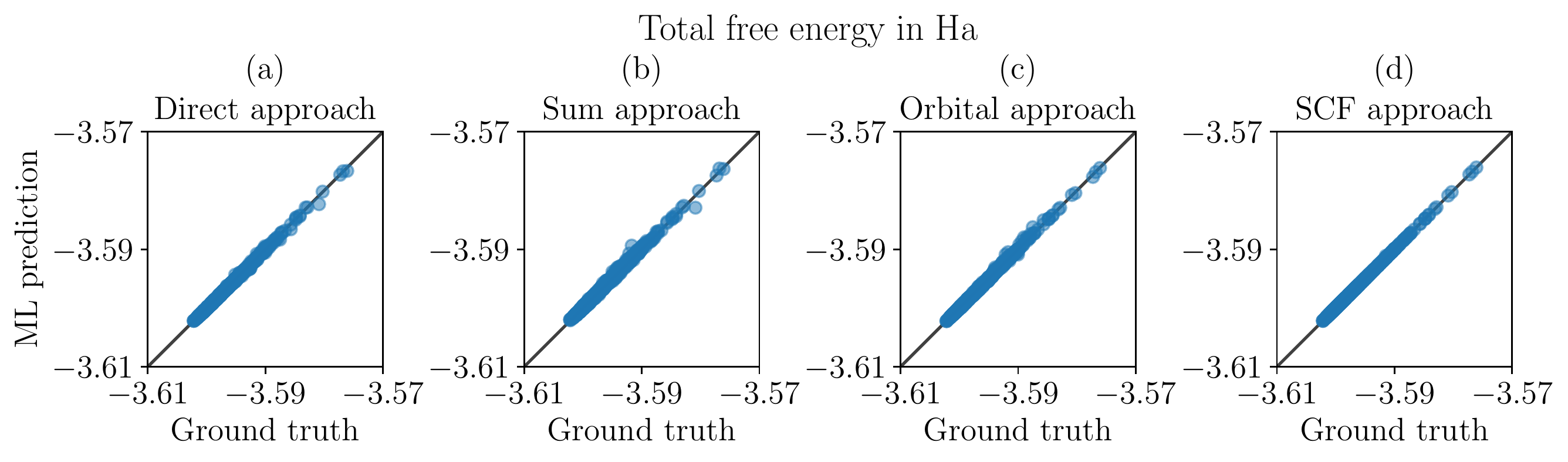}
	\end{center}
	\caption{Comparisons between predicted and true total free energy values for a 4-atom magnesium unit cell where the prediction is made using four separate ML approaches: (a) direct (same as Figure \ref{fig:prediction energy}(c)), (b) sum, (c) orbital, and (d) SCF.  The average absolute error changes from 0.10 mHa, 0.16 mHa, 0.087 mHa, to 0.0015 mHa, going from cases (a) to (d). Similarly, the maximum absolute error changes from 1.5 mHa, 2.3 mHa, 1.2 mHa, to 0.37 mHa.}
	\label{fig:free energy error}
\end{figure}

An intended goal of this work is the use of the electronic fields as pre-conditioners or predictor fields in larger multiscale calculations.  To evaluate their efficacy in doing so, we calculate the total free energy in four ways.  First, in the \emph{direct} approach, we learn the map from the deformation to the total energy ($\Phi_{\text{ml}}^{\mathcal{F}}$).  In the second \emph{sum} approach, we evaluate the total energy using equation (\ref{eq:free}) from the learned electron density, band structure energy and entropy fields, and atomic coordinates.  Third, in the \emph{orbital} approach, we use the learned electron density and atomic coordinates to construct the Hamiltonian, find the electronic orbitals, and then use the combination of the learned electron density and atomic coordinates with the computed orbitals (to compute the kinetic energy).  Note that in the third approach, we do not use the band structure energy, but effectively recompute it by computing the orbitals.  So, the difference between the direct and orbital energies is a measure of the inconsistency between the learned band structure energy and the learned energy density.   Finally, in the \emph{SCF} approach, we perform one SCF iteration starting from the learned electron density and atomic coordinates before computing the energy.  Note that we do not directly use the learned electron density, but recompute it using a SCF iteration.  So, the difference between orbital and SCF values indicates whether the learned electron density is close to convergence, or to use this electron density as a precursor.

Figure \ref{fig:free energy error} shows how closely the free energy predicted from the four approaches match to their true values. The sum approach leads to a very low average error of 0.15 mHa across all 1000 test data with the maximum error standing at 2.4 mHa. These errors are consistent with those of the band structure energy in Figure \ref{fig:prediction energy}, emphasizing the contribution of this band structure energy to the total  energy.  The orbital and SCF approaches lead to further reductions, and they even outperform the direct approach.  All of these signify that our proposed approach not only learns the energy but also the fields extremely accurately, and thus can be used as predictors in larger calculations.

\begin{figure}[h]
	\begin{center}
		\includegraphics[width=1.0\textwidth]{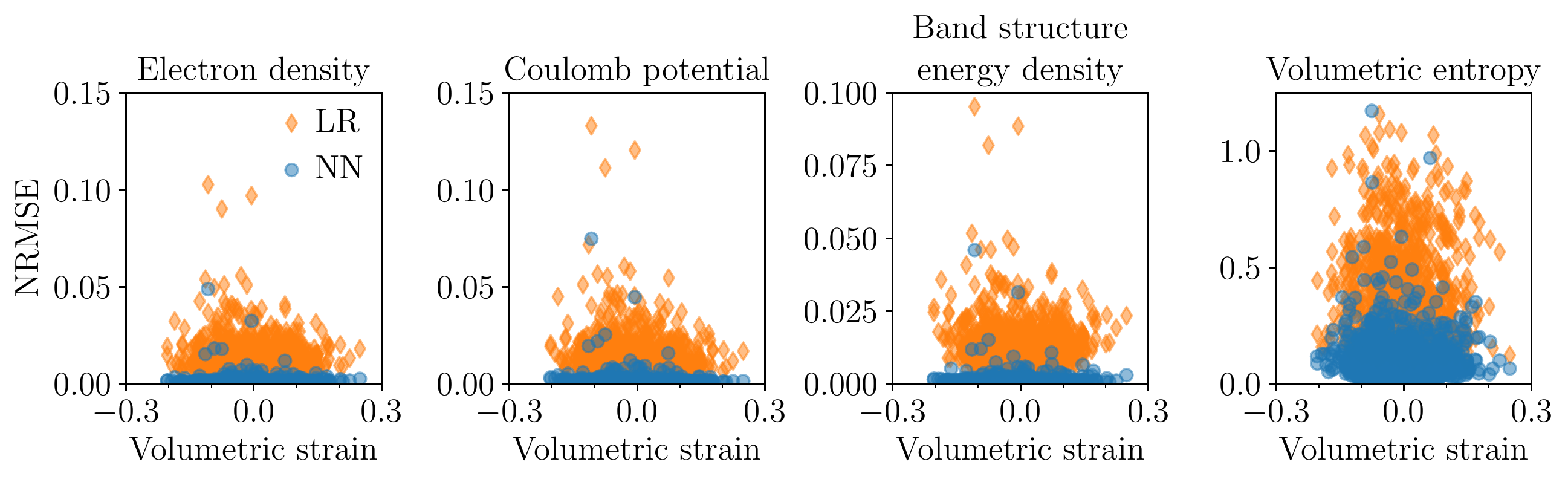}
	\end{center}
	\caption{Comparison of prediction errors when a linear regression (LR) is used in place of a neural network (NN) for each map $\Phi_{\text{ml}}$. The NN results are reproduced from Figure \ref{fig:error vs volumetric strain}.}
	\label{fig:error vs volumetric strain2}
\end{figure}

Finally, we compare our approximation with linear regression in Figure \ref{fig:error vs volumetric strain2}.  It shows that the error due to linear regression (LR) is significantly higher than that due to our nonlinear approximation using neural networks (NN), thus demonstrating the efficacy of our architecture (NN).

\subsection{Results: Stresses and instability}

In order to further assess the accuracy of the learned electronic fields and energy, and to understand its efficacy in practice, we study the derivative quantities (i.e. stresses) in the crystal subjected to deformation.   According to the Cauchy-Born rule \cite{e_mms_08}, the macroscopic deformation gradient in any macroscopic deformation is equal to the  matrix $F$ that maps the reference lattice to the current lattice.  The corresponding nonlinear strain measure $E =1/2 (F^TF - I)$ where $I$ is the identity.  The Cauchy or true stress in the material is given by 
\begin{equation}
\sigma = {1 \over V} F {\partial {\mathcal F} \over \partial E} F^T,
\end{equation}
where $V$ is the volume of the deformed unit cell and $ {\mathcal F}$ is the free energy of the unit cell.  It is common \cite{Nielsen1985,Ilawe2015} to approximate this using a small distortion approximation
\begin{equation}\label{eqn:stress-strain}
\sigma = \frac{1}{V(\varepsilon)} \frac{\partial \mathcal{F}}{\partial \varepsilon}
\quad \text{or} \quad 
\sigma_{ij} = \frac{1}{V(\varepsilon)} \frac{\partial \mathcal{F}}{\partial \varepsilon_{ij}},
\end{equation}
where $\varepsilon = 1/2 (F+F^T-I) \approx E$ is the linear strain measure.  We use this expression in our work, though the results can easily be adapted to the nonlinear counterpart.  The linear strain is related to the variables we use to describe the lattice ($\lambda_a,\lambda_b,\lambda_c,\theta_a,\theta_b,\theta_c$) through the relation
\begin{equation}\label{eqn:lambda to strain}
\lambda_a^2 = (1+\varepsilon_{11})^2 + \varepsilon_{12}^2 + \varepsilon_{13}^2, \quad
\lambda_b^2 = \varepsilon_{12}^2 + (1+\varepsilon_{22})^2 +  \varepsilon_{23}^2, \quad
\lambda_c^2 = \varepsilon_{13}^2 + \varepsilon_{23}^2 + (1+\varepsilon_{33})^2,
\end{equation}
\begin{equation}\label{eqn:theta to strain}
{\pi \over 2} - \theta_a = 2 \varepsilon_{23}, \quad 
{\pi \over 2} - \theta_b = 2 \varepsilon_{13}, \quad
{\pi \over 2} - \theta_c = 2 \varepsilon_{12},
\end{equation}
and we may obtain $\partial {\mathcal F}/ \partial \varepsilon_{ij}$ using the chain rule (see Appendix \ref{appendix:backpropagation} for details).

\begin{figure}[h]
	\begin{center}
		\includegraphics[width=0.9\textwidth]{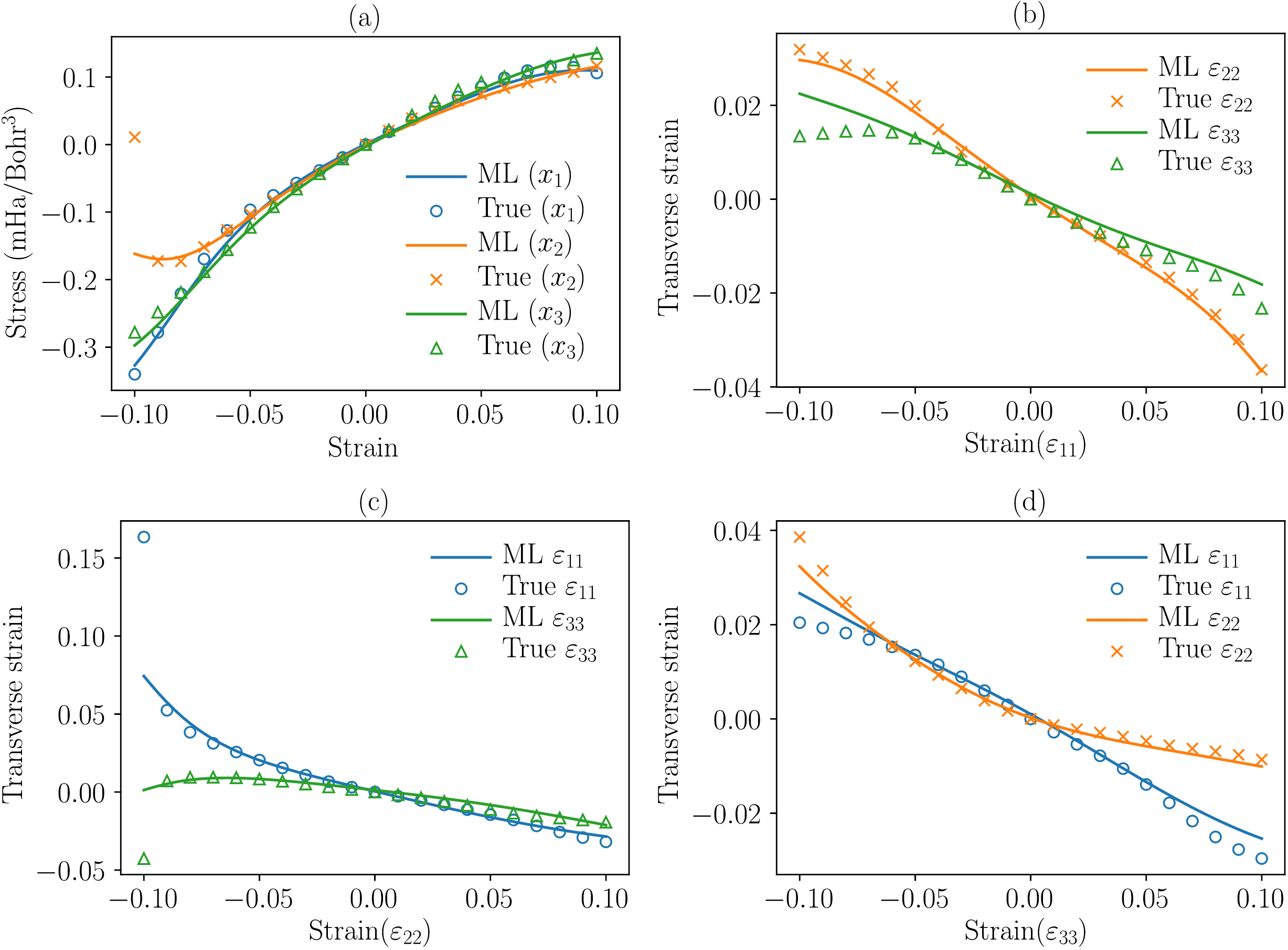}
	\end{center}
	\caption{Uniaxial test results. The results obtained from our machine learning model are marked as 'ML', while the results obtained directly from DFT calculations with stress relaxation are marked as 'true.' (a) Stress-strain curves for uniaxial stress in the $x_1$, $x_2$, and $x_3$ directions. (b--d) Corresponding transverse strain.}
	\label{fig:uniaxial stress-strain}
\end{figure}

\begin{figure}[h]
	\begin{center}
		\begin{subfigure}
			\centering
			\includegraphics[width=1.0\textwidth]{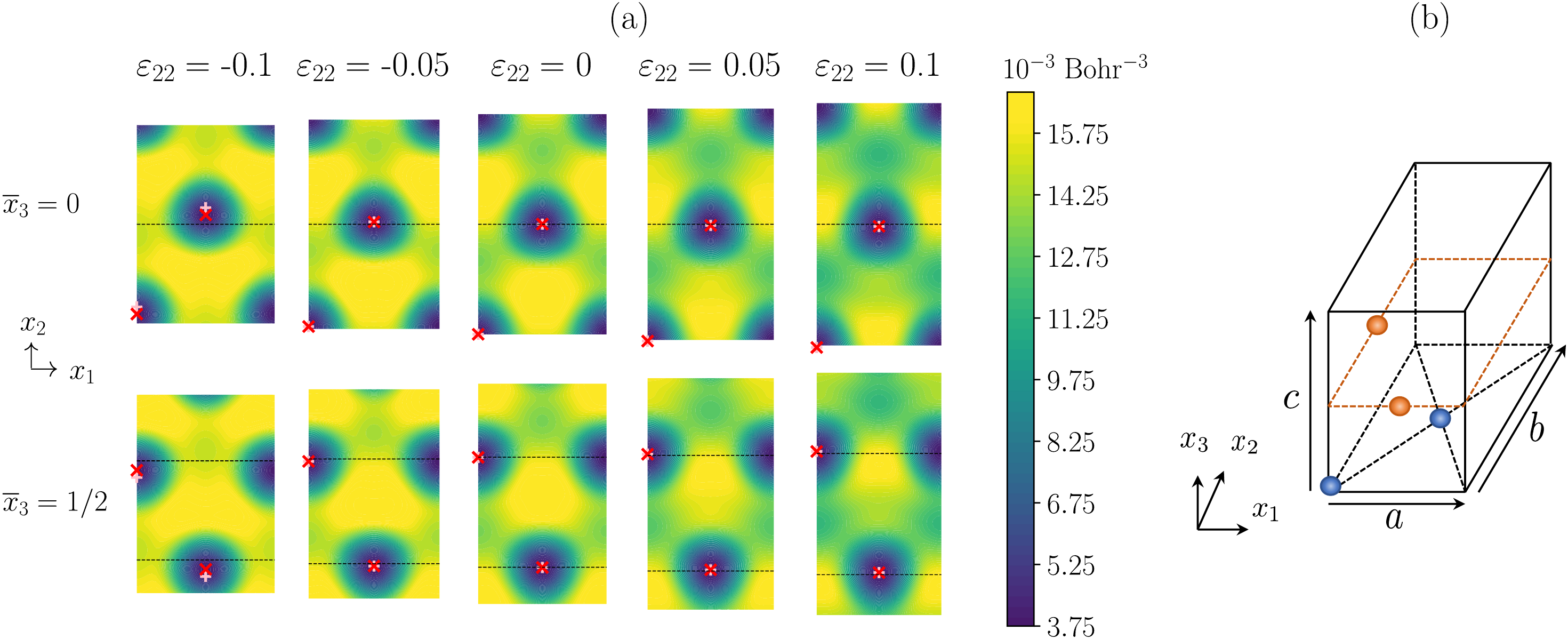}
		\end{subfigure}
	\end{center}
	\caption{Uniaxial test results in the $x_2$ direction obtained from our machine learning models. (a) Snapshots of a magnesium unit cell undergoing uniaxial test. The color gradient shows the electron density distribution, while the red crosses ($\times$) mark the atomic positions. The dashed lines on the plots correspond to $\bar{x}_2=1/2$ on the basal plane $\bar{x}_3=0$, as well as $\bar{x}_2=1/6$ and $\bar{x}_2=2/3$ on the basal plane $\bar{x}_3=1/2$. The actual atomic positions obtained from DFT calculations are shown as pink plusses ($+$) for reference. (b) An illustration of the deformed face-centered cubic (FCC) structure exhibited by the unit cell under uniaxial compression with $\varepsilon_{22}=-0.1$.}
	\label{fig:uniaxial x2}
\end{figure}

We focus specifically on the state of uniaxial stress where $\sigma_{ii}$ (for $i=$1, 2 or 3) is non-zero, but all the other components are zero.  We therefore prescribe $\varepsilon_{ii}$, and solve (\ref{eqn:stress-strain}) to obtain the other components of strain.   The solution to this equation is equivalent to minimizing the energy, and we use gradient descent with the step size chosen according to the Barzilai-Borwein method \cite{Barzilai1988}.  

The results are shown in Figure \ref{fig:uniaxial stress-strain} for the stress in the $x_1$, $x_2$, and $x_3$ directions as well as the corresponding transverse strains\footnote{ The use of cold smearing in our DFT calculations requires a pre-stress or residual stress to stabilize the HCP configuration.  We have subtracted the uniaxial residual stress in (a) and the corresponding residual strains in the other two transverse directions in (b--d) of this figure.}.  Figure \ref{fig:uniaxial stress-strain}(a) compares the ground truth stress to the ML approach, where the free energy is computed using the direct approach described earlier. We see that the ML approach predicts the stress extremely well except for high compression in the $x_2$ direction where there is an instability. Still, the approach captures the onset of this instability. Figures \ref{fig:uniaxial stress-strain}(b--d) show the corresponding transverse strains.  We see that this is not linear in any case.  Further, there is a dramatic expansion in the $x_1$ direction ($a$-axis) as we reach the instability during severe compression in the $x_2$ direction.

Figure \ref{fig:uniaxial x2}(a) shows the corresponding electronic fields and positions of the internal atoms.  Specifically, given the value of $\varepsilon_{22}$, we find the other components of strain and the corresponding deformation variables.  We then interrogate our learned maps to find the electronic fields and atomic positions.  The figure also compares the learned and ground truth atomic positions, and we see good agreement except at the instability. 
We observe that in most cases, the atoms in the unit cell lie very near to the dashed lines, thus exhibiting a deformed HCP structure. However the atomic positions change very abruptly under high compression in the $x_2$ direction. The learned atomic positions capture this abrupt change to some extent.

Together, Figures \ref{fig:uniaxial stress-strain} and \ref{fig:uniaxial x2} show that as the crystal is compressed in the $x_2$ direction and reaches its instability, it elongates dramatically along the $x_1$ direction.  Consequently, the lengths of $a$, $b$, and $c$ edges become similar.  This is accompanied by the relative sliding of the basal planes (see Figure \ref{fig:uniaxial x2}(a) where the atoms in the $\bar{x}_3=0$ plane move in the positive $x_2$ direction whereas the atoms in the $\bar{x}_3=1/2$ basal plane move in the negative $x_2$ direction), bringing the atoms on the middle basal plane to the face centers of the 4-atom unit cell.   Thus, the structure approaches that of a face-centered cubic (FCC) lattice as shown in Figure \ref{fig:uniaxial x2}(b).  In other words, we see a HCP to FCC phase transition.  This has been observed under high hydrostatic confinement \cite{lfz_jap_09}, but is generally overshadowed by the $(10\bar{1}2)$ tension twin mode at lower confinements\footnote{Magnesium has a soft ``tension'' twinning mode with a $(10\bar{1}2)$ twin plane and a $[\bar{1}011]$ shear direction \cite{cm_progmatsci_95}.   This twin causes a compression along the $[10\bar{1}0]$ or $x_2$ direction.  The observed instability is associated with this soft twinning mode.}.

\section{Conclusion}
In this paper, we have presented a machine learning approximation for the change in the electronic structure as a crystal is deformed.   We have demonstrated the approach on magnesium and shown that the machine-learned predictions reach the level of chemical accuracy. In particular, we not only  learn energy values accurately, but also predict electronic fields with minimal error. These show that the models can indeed be sufficiently accurate to be useful as predictors or pre-conditioners for large-scale DFT methods. Finally, we have computed derivative quantities such as stresses under specific loading conditions from the learning models and found that they match very well with the ground truth DFT results. The model can even capture the onset of strain-induced phase transformation. All these further indicate another future direction of extending the learning model to one that can extract DFT-informed constitutive relations that can be easily incorporated in continuum level calculations.

\section*{Acknowledgement}
We are grateful to the De Logi foundation and the Army Research Laboratory (under Cooperative Agreement Number W911NF-12-2-0022) for their generous support of the research.  The views and conclusions contained in this document are those of the authors and should not be interpreted as representing the official policies, either expressed or implied, of the Army Research Laboratory or the U.S. Government. The U.S. Government is authorized to reproduce and distribute reprints for Government purposes notwithstanding any copyright notation herein.

\begin{appendices}

\section{Band structure energy density, volumetric entropy, and total free energy in a crystal}
\label{appendix:derivation}

In the context of a crystal that has electronic orbital $\psi_{i,\bk}$ and energy state $E_{i,\bk}$ associated with each $\bk$ point in the Brillouin zone of the unit cell $\mathcal{U}$, the band structure energy density and volumetric generalized entropy are given by
\begin{equation}
u(\br) = \sum_{\bk \in \text{IBZ}} \sum_{i=1}^{N_{band}} w_{\bk} E_{i,\bk} \, f(E_{i,\bk}) |\psi_{i,\bk}(\br)|^2,
\end{equation}
\begin{equation}
s(\br) = -\frac{1}{\sqrt{\pi}} \sum_{\bk \in \text{IBZ}} \sum_{i=1}^{N_{band}} w_{\bk}  |\psi_{i,\bk}(\br)|^2 e^{-t^2} (\kappa t^3 + t^2 - \frac{1}{2}), \quad t = \beta(E_{i,\mathbf{k}}-E_f).
\end{equation}
Here we restrict $\bk$ to the irreducible Brillouin zone (IBZ) of $\mathcal{U}$ and add appropriate weights $w_{\bk}$ satisfying $\sum_{\bk \in \text{IBZ}} w_{\bk} = 1$. The energy eigenstates have also been normalized such that  
$ \int |\psi_{i,\bk}(\br)|^2 d\br = 1$ for all $i,\bk$.

Finally, we consider the total energy expression. The Kohn-Sham ground state energy is   
\begin{align}
E_{KS}[\{\psi_{i,\bk}\},\{R_I\}] =& \sum_{i=1}^{N_{band}} \sum_{\bk \in \text{IBZ}} -\frac{1}{2} w_{\bk} f(E_{i,\bk}) \int \psi_{i,\bk}^*(\br) \nabla^2 \psi_{i,\bk}(\br) d\br + E_{H}[\rho(\br)] + \int \rho(\br) V_\text{ext}(\br) d\br \nonumber \\
& + \int e_{xc}[\rho(\br)] d\br + E_\text{ion}[\{R_I\}] + E_{core},
\end{align}
where the first term of the expression is the independent-particle kinetic energy, \\ \noindent $\rho(\br) = \sum_{i=1}^{N_{band}} \sum_{\bk \in \text{IBZ}} w_{\bk} f(E_{i,\bk}) |\psi_{i,\bk}|^2$ is the valence electronic density, $E_\text{H}$ is the Hartree energy (or in other words, the classical Coulomb interaction energy of the electron density interacting with itself), and $E_{ion}$ is the Coulomb energy associated with interactions among the ions at positions $\{R_I\}$ which is computed using the Ewald summation method. There is also an additional energy contribution known as $E_{core}$ arising from the fact that the ion is not a point charge. Its exact expression can be found in \cite{Payne1992}.

Using the Kohn-Sham equation $\mathcal{H}_{KS} \psi_{i,\bk} = E_{i,\bk} \psi_{i,\bk}$ with $\mathcal{H}_{KS} = -\frac{1}{2} \nabla^2 + V_\text{H} + V_\text{ext} + V_\text{xc}$ and orthonormality condition $\int \psi_{i,\bk}^*(\br) \psi_{i,\bk}(\br) d\br = \delta_{i,i'} \delta_{\bk,\bk'}$, we may rewrite the ground state energy in terms of the band structure energy $U$ analogous to equation (\ref{eq:free}) as follows:
\begin{align}\label{eqn:int energy}
E_{KS}[\{\psi_{i,\bk}\},\{R_I\}] =& \underbrace{ \sum_{i=1}^{N_{band}} \sum_{\bk \in \text{IBZ}} w_{\bk} f(E_{i,\bk}) E_{i,\bk}}_{U} 
- E_{H}[\rho(\br)] 
- \int V_{xc}[\rho(\br)] \rho(\br) 
+ \int e_{xc}[\rho(\br)] d\br \nonumber \\
&+ E_{ion}[\{R_I\}] 
+ E_{core}.
\end{align}

Then, the total Helmholtz free energy of the system is simply
\begin{equation}\label{eqn:free energy}
\mathcal{F} = E_{KS} - S/\beta.
\end{equation}

\section{Backpropagation method in obtaining energy derivatives}
\label{appendix:backpropagation}
The machine learning model that maps deformation $D=\{\lambda_a,\lambda_b,\lambda_c,\theta_a,\theta_b,\theta_c\}$ to the total free energy $\mathcal{F}$ has the following mathematical structure:
$$
\mathbf{h}^{(0)} = {\mathbf{W}^{(0)}}^{-1} \left( \mathbf{h}^{(-1)} - \mathbf{b}^{(0)} \right), \quad \mathbf{h}^{(-1)}=[\lambda_a \; \lambda_b \; \lambda_c \; \theta_a \; \theta_b \; \theta_c]^T, \quad
W_{ij}^{(0)} = 0 \text{ if } i\neq j,
$$
$$
\mathbf{h}^{(k)} = g \left( {\mathbf{W}^{(k)}}^T \mathbf{h}^{k-1} + \mathbf{b}^{(k)} \right), \quad k = 1,2,3,
$$
$$
h^{(4)} = {\mathbf{W}^{(4)}}^T \mathbf{h}^{(3)} + b^{(4)},
$$
$$
\mathcal{F} = h^{(5)} = W^{(5)} h^{(4)}  + b^{(5)}.
$$
The first and the fourth lines represent the preprocessing steps that remove the mean and variance from the data, where $\mathbf{b}^{(0)},\mathbf{W}^{(0)},\mathbf{b}^{(5)},\mathbf{W}^{(5)}$ are the mean and standard deviation of the input $D$ and output $\mathcal{F}$ of the model. The second and third lines indicate the dense neural network with three hidden layers, hyperbolic tangent activation function $g(x) = \tanh(x)$, and fitted weights $\mathbf{b}^{(k)},\mathbf{W}^{(k)}$ with $k=1,2,3$. We note that we have intentionally chosen an activation function with smooth derivative which is the case of hyperbolic tangent function to allow for easy minimization of energy $\mathcal{F}$ via the gradient descent method.

The derivatives of $\mathcal{F}$ with respect to $\mathbf{h}^{(-1)}$ can be computed as follows:
$$
\frac{\partial \mathcal{F}}{\partial h_i^{(3)}} = W^{(5)} W_{i1}^{(4)},
$$
$$
\frac{\partial \mathcal{F}}{\partial h_i^{(k-1)}} = \sum_j \frac{\partial \mathcal{F}}{\partial h_j^{(k)}} g'(({\mathbf{W}^{(k)}}^T \mathbf{h}^{(k-1)}+\mathbf{b}^{(k)})_j) W_{ij}^{(k)}, \quad k = 3,2,1,
$$
$$
\frac{\partial \mathcal{F}}{\partial h_i^{(-1)}} = \frac{\partial \mathcal{F}}{\partial h_i^{(0)}} \frac{1}{W_{ii}^{(0)}}.
$$

Subsequently, applying the chain rule gives
\begin{equation*}
\frac{\partial \mathcal{F}}{\partial \varepsilon_{kl}} = \sum_{i=1}^{6} \frac{\partial \mathcal{F}}{\partial h_i^{(-1)}} \frac{\partial h_i^{(-1)}}{\partial \varepsilon_{kl}}.
\end{equation*}

\end{appendices}

\bibliographystyle{unsrt}
\bibliography{References}

\end{document}